
\input phyzzx
\def\bracket#1#2#3{\Big\langle{#1}\Big\vert{#2}\Big\vert{#3}\Big\rangle}
\def\braket#1#2{\left\langle #1  \vert  #2\right\rangle}
\def\nint{\int\nolimits}
\def\noint{\oint\nolimits}

\titlepage
\title{ Geometric Phases, Coherent States and Resonant Hamiltonians}
\author { Masao Matsumoto}
\address{ 1-12-32 Kuzuha Asahi, Hirakata City, Osaka 573, JAPAN}
\author { Hiroshi Kuratsuji}
\address{Department of Physics,
Ritsumeikan University-BKC, Kusatsu City 525-77, JAPAN}

\abstract{
We study characteristic aspects of the geometric
phase which is associated with the generalized coherent states.
This is determined by special orbits in the parameter space
defining the coherent state, which is obtained as a solution of the
variational equation governed by a simple model Hamiltonian
called the "resonant Hamiltonian".  Three typical coherent states
are considered: SU(2), SU(1,1) and Heisenberg-Weyl. A possible experimental
detection of the phases is proposed in such a way that
the geometric phases can be discriminated from the dynamical phase.}

\endpage

\chapter{Introduction}
The main feature of quantum mechanics is the exsistence
of the probability amplitude which underlies all the atomic processes.
\Ref\Dirac{P.~A.~M.~Dirac, ~Fields and ~Quanta {\bf 3}(1972)139.}
In particular, the phase factor of probability amplitude
has recently renewed interest, which is inspired by a specific
motivation. Namely, if one considers the cyclic change in
quantum system, one gets the so-called geometric phases.
\Ref\Shapere{
A.~Shapere and ~F.~Wilzcek eds.
"~Geometric ~Phases in ~Physics, ~World ~Scientific (1989).}
 From the historical point of view, Dirac was the first who recognized
the geometric phase in the form of a non-integrable phase
(or path dependent phase factor).
\Ref\DiracII{P.A.M.~Dirac, Proc. Roy. Soc.{\bf A133}(1931)60.}
Apart from this monumental work,  the development of
the geometric phases has, roughly speaking, two aspects.
One aspect is that the origin of geometric phases dates back
to the Bohr-Sommerfeld factor $ \exp[i\oint pdx] $
which is closely connected with the symplectic (canonical) structure
of the classical phase space.
 From the modern point of view, this original form of the geometric
phases has been further extended to the one defined over the generalized
(or curved) phase space.
\Ref\KuraI{H.~Kuratsuji, ~Phys.~Rev.~Lett.{\bf 61}(1988)1687
and references cited therein.}
The present paper is mainly motivated by this aspect to
obtain much deeper understanding of the
characteristic features of the geometric
phases associated with the generalized phase space.
The other aspect is that the geometric phase in its literal sense
has been known since early sixties in connection with molecular physics
\Ref\Hertzberg{G.~Hertzberg and ~H.~C.~Longuet-Higgins
{}~Dics.~Farad.~Soc. {\bf 35}(1963)77;
A.J.Stone, Proc.Roy.Soc.{\bf 351}(1976)141.}
Then one has been led to the final goal of the nowadays
famous quantum adiabatic phase; alias "Berry's phase", which
has been formulated by the conventional Schr$\ddot{\rm o}$dinger
equation
\REF\Berry{M. Berry, Proc.Roy.Soc.{\bf 392}(1984)45}
\REF\Simon{B.Simon, Phys.Rev.Lett.{\bf 51}(1983)2167.}
\refmark{\Berry-\Simon}
or path integral method.
\Ref\Iida{H.Kuratsuji and S.Iida, Prog.Theor.Phys, Phys.
{\bf 74}(1985)439.}
The contributions in the early development of this topics
as well as the historical ones may be found in
the reprint book edited by Shapere and Wilczek.
\refmark{\Shapere}

What we want to address here is to investigate some specific aspect
of the geometric phases formulated by path integral
in terms of the so-called generalized coherent state, which
has originally been studied for different purpose,
(see, e.g. the paper
\Ref\KS{
 H.~Kuratsuji and T.~Suzuki, J. Math. Phys.{\bf 21}(1980)472.}
) and later has been restudied for the purpose of describing
the geometric phases without recourse to the adiabaticity.
\refmark{\KuraI}
Although the utility of this aspect has not yet been widely appreciated,
we believe that it would commit to crucial points of physics
concerning the geometric phase problems. The generalized coherent
state (GCS) denoted by $\ket{Z}$ is parametrized by a point  $Z$
on some manifolds (mostly complex ones)
\Ref\Per{A.~M.~Perelomov,"~Generalized ~Coherent ~State",
(Springer, Berlin,1985)}
whose coordinates may be linked to the external or
macroscopic parameters characterizing the systems.
By considering the propagator for the roundtrip along
a closed loop in the generalized phase space,
the geometric phase is written as a contour integral of
the connection factor that is given as the overlap function between
two nearby coherent states.
As will be explained in Section 2, the condition determining
the geometric phase is given by the variation equation leading
to the paths (or orbits) ~in the parameter space or the generalized
phase space. The variation principle plays a role of a substitute
for the adiabatic condition determining the adiabatic phases.

In this paper we do not intend to develop a general formalism
of the geometric phase, but aim to examine rather specific cases
relevent to direct observation by experiment. Namely, we are concerned
with the particular cases such that the path is determined by the
special class of model Hamiltonians called the resonant Hamiltonian
which is given in terms of the linear "generalized spin"
in an oscillating "magnetic field".
\Ref\Matsu{
H.Kuratsuji and M.Matsumoto; Phys.Lett.{\bf A155}(1991)99.}
These model Hamiltonian may be realized in quantum optics or
similar device. The generalized spins
form the generators of specific Lie groups
which are connected with the GCS under consideration.
For this particular class of coherent states, the geometric
phases are calculated by using the specific solutions
called the resonance solutions that are derived by the resonant
Hamiltonian. In the following we consider three typical GCS's of
compact and non-compact types, namely, the spin (or SU(2)) CS,
and the Lorentz(or SU(1,1))CS and the boson CS.
We show that a detection is possible for
the geometric phases accompanying the resonant solutions corresponding
to these three types of coherent states.

\chapter{ Preliminary}
\par
We start with a concise review of the general theory of the
geometric phase formulated in terms of the GCS.
\refmark{\KuraI}
Consider the propagator starting from and ending
at the state$\ket{Z_0}$ during the time interval $ T $:
$$
K = \bra{Z_0} P  \exp{-{i \over \hbar}\nint_0^T{\hat H}(t)\,dt}
\ket{Z_0}\eqn\Aone
$$
where the Hamiltonian is time-dependent,
which means that the evolution operator generally
reads time-ordered product. \Aone~represents the probability
amplitude for coincidence; the amplitude for
a cyclic change that the system starts with the state $ \ket{Z_0} $
and returns to the same state after a time interval ~$ T $.
That implies that the system proceeds along closed paths in the Hilbert
space spanned by the set of GCS.  Using the partition of unity at
each infinitesimal time-interval, we have
$$
K = \int \prod_{k=1}^{\infty}\braket{Z_{k-1}}{Z_k}
\exp[-{i\over \hbar}\nint_C \bracket{Z}{\hat H}{Z}dt]D\mu(Z)
\eqn\Atwo
$$
where  $ D\mu(Z) \equiv \prod_{t=0}^T d\mu(Z(t))$  and
$ d\mu(Z) $ denotes the invariant measure on the
generalized phase space specified by the complex vector
$ Z=(z_1,z_2,\cdots,z_n) $. In \Atwo, the infinite product
represents the finite connection along the closed loop in the
complex parmeter space, in which each infinitesimal factor
represents the connection between two infinitesimally separated points.
If use is made of the approximation
$ \braket{Z_{k-1}}{Z_k} \simeq \exp[i\bracket{Z}
{\partial \over \partial t}{Z}dt] $, \Atwo~ is written as
the functional integral over all closed paths
$$ K = \int \exp[{i \over \hbar}\Phi(C)]D\mu(Z),
\eqn\Atwo'
$$
where $ \phi $ is nothing the "action functional":
$$ \Phi(C)
= \nint_0^T \bra{Z} i \hbar {\partial \over \partial t}
-\hat H(t) \ket{Z} dt ,
\eqn\Athree
$$
where $ H(Z,Z^*,t) \equiv \bra{Z} \hat H \ket{Z}\equiv H(t) $.
Specifically, we write the first term of $S$ as
$$
\Gamma(C)=\noint_C \bra{Z} i \hbar {\partial \over \partial t}
\ket{Z} dt,       \eqn\Afour
$$
which give nothing but the geometric phase in terms of the GCS,
which is quoted simply as $\Gamma$ hereafter.
On the other hand, the second term of \Athree~is called
the Hamiltonian term denoted as $ \Delta $:
$\Delta(C)=\oint_C \bracket{Z}{\hat H}{Z} dt$.
If one uses the kernel function
$$ F(Z,Z^*)=\braket{\tilde Z}{\tilde Z}
$$
($\ket{\tilde Z}$ being the unnormalized CS),
$ \Gamma $ is cast into the form
$$    \Gamma = \noint \sum_{k=1}^n {i \hbar \over 2}
\left({\partial \log F \over \partial z_k} dz_k
-{\partial \log F \over \partial {z_k}^*} d{z_k}^*\right)
\equiv \oint \omega \eqn\Afive
$$
In order to calculate the explicit form of $ \Gamma $,
we need to select a specific cyclic path $ C(Z(t)) $ in the
generalized phase space.
This may be realized by considering the semiclassical limit of
\Athree; the stationary phase condition
$ \delta S= 0 $ yielding the equations of motion for $ Z $
$$
i \hbar \sum_{j=1}^n g_{i \bar j}
\dot z_j={\partial  H\over \partial z_i^*},
-i\hbar \sum_{j=1}^n g_{i \bar j} \dot z_j^*
={\partial H \over \partial z_i}\eqn\Aseven
$$
where
$$
g_{i \bar j}
={\partial^2 \log F \over \partial z_i \partial z_j^{*}},\eqn\eight
$$
which denotes the metric of the generalized phase space:
the so-called Kaehher metric. The propagator is thus reduced to
a simple form
$$        K_{sc} = \exp[{i\Gamma(C) \over \hbar}]
\exp[-{i\Delta\over \hbar}]
\eqn\Anine
$$
Namely, if there exist closed paths,
the propagator may be expressed as the overlap between
two coherent states
$$
K_{sc} = \braket{Z_0(T)}{Z_0(0)}\eqn\Aten
$$
where the ket vector is parametrized by the orbit,
the end point of which coincides with $ Z_0 $ at the time T.
In this way the final state may accomodate the history
which the system develops.

To consider the semiclassical limit should be compared with
the procedure adopted in getting the adiabatic quantum phase,
where the change of state vector is governed by the cyclic motion
that evolves adiabatically.
\refmark{\Iida}
In the present case, the adiabaticity is not
necessary and the principle governing the geometric phase
is played by the
"quantum variational principle" leading to the equation
of motion in the parameter space that defines the generalized
coherent state.   Some explanation is in order regarding the meaning of
the choice of the closed path.  Taking the semiclassical limit
means that the condition is fixed for choosing the specific
path among all the possible paths, where the parameter controlling
the semiclassical limit is played by the Planck constant.
This feature partially corresponds to the situation of
taking the adiabatic limit for the quantum transition
for which the transition takes place between the states
labelled by the discrete eigenstates possessing with
the same quantum number.

Here a comment is given for a possible experimental
detection of the phase $ \Gamma $.
As is suggested in the above,
after a cyclic motion the state vector totally aquires the phase,
which by eq. \Athree~ consists of two parts:
$ \Phi={(\Gamma - \Delta) \over \hbar} $:
the geometric and the Hamiltonian term.
In this point it is crucial to separate these two terms
from each other in actual situation;
especially, to extract the geometric term $ \Gamma $.
In this connection, the problem is the fact that
it is not easy in general to obtain the phase in a concise manner.
The reason arises from the difficulty of finding out the cyclic
path relevant to evaluate the phase $\Gamma(C)$.
So we must resort to the special situation that
enables us to extract the cyclic path in a simple way.
In the next section we shall realize this program.

\chapter{The Geometric Phases for Three Types of Coherent States}

In this section we consider the three typical coherent states
to demonstrate the specific feature of the geometric phase $ \Gamma $.

\section{The case of SU(2) coherent state}

As a first example we examine the case of a spin system
which is described by the SU(2) coherent state.
Consider a particle with spin J in a time-dependent
magnetic field, which has the component:
$$
{\bf B}(t)=(B_0 \cos \omega t,B_0 \sin \omega t,B)\eqn\Bone
$$
namely, a static field along the z-axis plus a time-dependent
field rotating perpendicular to it with the frequency $\omega$,
which is familiar in magnetic resonance.
The system may be described by the spin or SU(2)CS
(alias Bloch state): $\ket{z}$ is defined as
$$
\ket{z}=(1+\vert z \vert^2)^{-J} e^{z \hat J_+} \ket{0} ,
\eqn\Btwo
$$
where $\ket{0}=\ket{J,-J}$ satisfying
$\hat J_z\ket{0}=-J\ket{0}$ and $\hat J_{\pm}$ are usual spin operators
and z takes any complex values.  The Hamiltonian of the system is
$$
\hat H(t) =-\mu \bf B (t) {\scriptstyle{\bullet}} \bf J  ,\eqn\Bthree
$$
where ${\bf J} \equiv (\hat J_1,\hat J_2,\hat J_3)$
is a matrix vector satisfying
$ {\bf J} \times {\bf J} = i \bf J $.
Note that the state is specified by a single complex parameter
experessed in terms of the polar coordinate
$$
z=\tan {\theta \over 2} e^{-i \phi}~(0 \le \theta \le \pi ,
0 \le \phi \le 2 \pi )\eqn\Bfour
$$
This suggests that the generalized phase space
is isomorphic to $S^2$ in the case of the spin CS .
The expectation value of the Hamiltonian
$H(t)=H(z,z^*,t)$ is thus given as
$$
H(z,z^*)=H(\theta,\phi)
= - \mu J [B_0 \sin \theta \cos (\phi-\omega t)-B \cos \theta]
\eqn\Bfive
$$
Now let us write the variation equation for the case of spin CS.
By using the kernel function $ F=\braket{ \tilde z} { \tilde z }
= (1+\vert z \vert^2)^{2J} $, we get
$$
2iJ\hbar {\partial  z \over \partial t}
=(1+\vert z \vert^2)^2 {\partial \it H \over \partial z^*},\eqn\Bsix
$$
    together with its complex conjugate. This is alternatively
written in terms of the polar coodinate,
$$
\dot \theta
=-{1 \over J \hbar \sin \theta}{\partial H \over \partial \phi} ,
\dot \phi
={1 \over J \hbar \sin \theta }
{\partial H \over \partial \theta } .\eqn\Bseven
$$
For the case of \Bfive, it turns out to be
$$
\dot\theta
= -{ \mu B_0 \over \hbar} \sin (\phi - \omega t ),
\dot \phi
= - {\mu \over \hbar}[ B_0 \cot \theta \cos ( \phi - \omega t ) + B ]     ,
\eqn\Beight
$$
One sees that this form of equations of motion allows
a special solution
$$
\phi = \omega t, \theta = \theta_0 (= {\rm const} ),\eqn\Bnine
$$
where the following relation should hold among
the parameters $ \theta_0, B, B_0$:
$$
\cot \theta_0
= - \left( { B \over B_0 }+ {{\hbar \omega}
\over \mu B_0}\right).\eqn\Bten
$$
The solution of the form \Bnine~may  be called the
"resonance" solution, since it corresponds to the
one for the case of the forced oscillation.
The set of parameters $(B,B_0,\omega)$ satisfying \Bten~for a
fixed value $\theta = \theta_0$ belong to a family of
resonance solutions. Indeed, this set of parameters
forms a surface in the paramter space $(B,B_0,\omega)$,
which we call \lq\lq invariant surface" hereafter and
characterizes the resonance condition.
Equation \Bnine~ gives a definite cyclic trajectory
$(\theta,\phi)$ with the period $T = 2\pi/\omega$ in the
generalized phase space.
The condition \Bten~is crucial, since
the quantities in the right hand side of \Bten~ are
all given in terms of constants that may be allowed
to be compared with experiment.

Next we turn to the evaluation of the phase
$ \Gamma $ that is fitted with this special solution.
 From \Btwo~ one gets
\refmark{\Iida}
$$
\eqalign{
\Gamma(C) & = \oint \bracket{z}{i\hbar
{\partial \over \partial t}}{z}dt \cr
          & = \oint{iJ\hbar \over {1+\vert z \vert^2}}
          (z^{*}\dot z -c.c)dt \cr
          & = \oint J\hbar(1-\cos\theta)d\phi }
\eqn\Beleven
$$
By noting the resonance solution, this becomes
$$
\Gamma(C)=2\pi J \hbar (1-\cos \theta_0)=-J \hbar \Omega(C),
\eqn\Btwelve
$$
$ \Omega(C) $ is nothing but the solid angle subtended
by the curve C at the origin of the phase space, which was
first used for the case of the adiabatic phase.
On the other hand, the Hamiltonian phase  $\Delta $ is given by
$$
\Delta(C)={ 2 \pi \mu J \over \omega }
( B_0 \sin \theta_0 - B \cos \theta_0 ).
\eqn\Bthrteen
$$
The important point is that the phase $ \Gamma $
depends only on~$\theta_0$.
Therefore any point lying on the "invariant surface"
gives the same $ \Gamma $.
This fact may play a crucial role for extracting the geometric part
$ \Gamma $ from the total phase that can be detected
in possible experimental situations. The detail will be discussed in the
next section.  On the other hand, the Hamiltonian phase is not
determined solely by $\theta$.

\section{The case of Lorentz coherent state}

In this subsection we consider the phase $ \Gamma $
that is connected with a non-compact coherent state;
the SU(1,1) CS; (or alternatively called
the Lorentz coherent state),
since SU(1,1) is locally isomorphic to Lorentz group
of 2+1 dimension.  First we give a brief explanation for
the generators of Lorentz CS. The algebra we need here is given
by a set of bilinear forms of boson creation and anihilation operators
for a single mode electromagnetic field:
$$
\hat K_{+} = {1 \over 2} ({\hat a}^{\dagger})^2,
{}~\hat K_{-} = {1 \over 2} {\hat a}^2,
{}~\hat K_0 = {1 \over 4} (\hat a {\hat a}^{\dagger}
+{\hat a}^{\dagger} \hat a).\eqn\Lone
$$
It is known that the discrete series of the irreducible representation
of SU(1,1) is divided into two classes characterized by the number
$ k={1 \over 4} $ or $ k={3 \over 4} $.
We see that for these two cases $ \ket{0} $ represents the state of
the photon number zero (namely, the vacuum state)
and the state of photon number one respectively.
Using this realization, we have the so-called
"squeezing operator"
$$
S(\zeta)= e^{\zeta \hat K_+ - \zeta^* \hat K_-}
= e^{{1 \over 2} ( \zeta ({\hat a}^{ \dagger})^2
- \zeta^* {\hat a}^2 )}\eqn\Ltwo
$$
with a squeezing parameter $ \tanh \vert \zeta \vert $
and a rotating angle $\phi/2$.
\Ref\Caves{
See e.g. C.~M.~Caves,~Phys.~Rev.~{\bf D23}~(1981)1693.}
By applying $S(\zeta)$ to the vacuum state, $
\ket{0} \equiv \ket{k,m=0} $,
we have the Lorentz CS:
$$
\ket{z}=e^{\zeta \hat K_+ - \zeta^* \hat K_-} \ket{0}
= (1-\vert z \vert^2)^k e^{z \hat K_+} \ket{0},
\eqn\LtwoX
$$
By using this form, the phase $ \Gamma $ for the Lorentz CS
is calculated by following a manner similar to the case of SU(2) CS.
In terms of the complex representation, it is given by
$$
\Gamma(C) = -i\hbar k \noint_C{z \dot z^* -\dot z z^*
\over 1-\vert z \vert^2} dt.
\eqn\Lthree
$$
or using the angle parameters
$$
z= \tanh({\tau \over 2})e^{-i\phi},\eqn\Lfour
$$
we have
$$
\Gamma(C) = \int\hbar k (\cosh \tau - 1)\dot\phi dt.
\eqn\Lfive
$$
The Lagrangian is given by
$$L(\tau,\phi)
= \hbar k (\cosh \tau - 1) \dot \phi
- H(\tau,\phi)
\eqn\Lsix
$$
and hence the variation equation leads to
$$
\dot\phi
= {1 \over \hbar k \sinh \tau} {\partial H \over \partial \tau},
\dot\tau
= -{1 \over \hbar k \sinh \tau} {\partial H \over \partial \phi}.
\eqn\Lseven
$$
Now, consider the system which is composed of cavity mode and
the squeezed state generating interaction.
\Ref\Walls{
P.~D.~Drummond and D. F. Walls, J. Phys. A {\bf 13}(1980),725.}
The Hamiltonian is given by
$$
\hat H = \hbar \omega_0 ({\hat a}^{ \dagger} \hat a + {1 \over 2})
+ \hbar[V^* {({\hat a}^{\dagger})}^2 + V {\hat a}^2 ],
\eqn\Leight
$$
where $ V $ is the interaction parameter including the effect
of pumping light.  Here we take $ V = \kappa e^{-i \omega t} $
: an oscillating pumping light.  In terms of the pseudo-spin,
the Hamiltonian can be expressed as
$$\eqalign{ \hat H
& = 2\hbar [ \omega_0 \hat K_0 +\kappa ( e^{i \omega t} \hat K_+
+ e^{-i \omega t} \hat K_-)]\cr
&  \equiv  {\bf C}~ \tilde {\scriptstyle \bullet} ~ {\bf K}, \cr}
\eqn\Lnine
$$
where $ {\bf C}~ \tilde {\scriptstyle \bullet} ~ {\bf K}
\equiv C_0 \hat K_0 - C_1 \hat K_1 - C_2 \hat K_2 $.
Here the magnetic field analogue ("pseudo-magnetic field", say)
is given by
$$
{\bf C} = (C_0,C_1,C_2)
=2\hbar (\omega_0 , - 2 \kappa \cos \omega t,
- 2 \kappa \sin \omega t ),
\eqn\Lten
$$
The expectation value of $ H $ becomes
$$
H(t) \equiv \bra{z} \hat H \ket{z}
=  H (\tau,\theta)
=  2 \hbar k[\omega_0 \cosh \tau +
2 \kappa \sinh \tau \cos(\phi - \omega t)].
\eqn\Leleven
$$
In this way, we get the equation of motion in terms of
the angle variables:
$$
\dot \phi = 2[ \omega_0 + \kappa\coth\tau\cos(\phi - \omega t)],
{}~\dot \tau = 4 \kappa \sin(\phi - \omega t ),\eqn\Ltwelve
$$
We obtain a \lq\lq resonance solution" in a completely similar
manner to the previous SU(2) case;
$$
\phi =  \omega t,
\tau = \tau_0 (= const),
\eqn\Lthirteen
$$
provided the following relation is satisfied:
$$
\coth \tau_0 =
{-\omega - 2 \omega_0 \over 4 \kappa},\eqn\Lfourteen
$$
where
$
(\omega, \omega_0, \kappa)
$
should obey the condition
$$
\vert{\omega + 2 \omega_0 \over 4 \kappa} \vert >1,
$$
since $ \vert \coth x \vert >1 $ .
Note that the orbit given by \Lthirteen~forms a circle on the
"pseudo-sphere". The condition \Lfourteen~ just determines the
"invariant surface "~in the parameter space
($\omega$, $\omega_0$, $\kappa$) on which $\tau_0$ is constant.
For the path C described by this solution, the phase $ \Gamma $ becomes
a simple form
$$
\Gamma(C) = 2 \pi  \hbar k (\cosh \tau_0 - 1 )
\eqn\Lfifteen
$$
and the Hamiltonian phase is given by
$$
\Delta(C) = {4 \pi \hbar k \over \omega}
( \omega_0 \cosh \tau_0 + \kappa \sinh \tau_0 ).\eqn\Lsixteen
$$
We arrive at a result that is completely paralel with the case
of spin CS: the geometric phase depends only on the \lq\lq
invariant surface" and the Hamiltonian phase
does not satify this property.

\section{ The case of boson coherent state}

 As a third example, we consider a harmonic oscillator
driven by an external force for which the Hamiltonian is given by
$$
\hat H
= {1 \over 2} ( \hat p^2 + {\omega_0}^2 \hat q^2)
+F(t)\hat q,
\eqn\Cone
$$
This can be written by boson creation and anhialation operators as
$$
\hat H
=\hbar \omega_0 {\hat a}^{\dagger} \hat a
+\beta(t) {\hat a}^{\dagger} +\beta^*(t)
\hat a.
\eqn\Ctwo
$$
This type of Hamiltonians appears in the problems of detecting
gravitational radiation
\REF\Hollen{
J.~N.~Hollenhorst, Phys. Rev. D {\bf 19}(1979)1669.}
and/or quantum optics.
\REF\Loudon{
R.~Loudon,~"Quantum Theory of Light"~( Pergamon press, \hfill\break
Oxford, ~1973 )}
\refmark{\Hollen, \Loudon}
In the following we shall take up the second one, for example,
and discuss the possibility of finding the effect of
the geometric phase.  Consider a single mode electric field
inside a cavity driven  externally by a coherent driving field.
If we neglect the cavity damping,  we have the Hamiltonian:
$$
\hat H = \hbar \omega_0 {\hat a}^{\dagger} \hat a
     + \hbar ({\hat a}^{\dagger} E(t) e^{-i\omega t}
     + \hat a E^*(t) e^{i \omega t}),
\eqn\Cthree
$$
which belongs to the type of \Ctwo. The first term represents the
cavity mode Hamiltonian, where $ \omega_0 $ means the fundamental
cavity resonance and the second term gives the Hamiltonian for the
coherent driving field respectively.  Here $ E(t)$ is the driving
field amplitude, while $\omega$ means the driving frequency.
The coherent state is now given by a standard (boson) coherent state:
$$
\ket{z}
=e^{-{1\over2}\vert z \vert^2} e^{z {\hat a}^{\dagger}} \ket{0}
\eqn\Cfour
$$
where the relation $  \hat a \ket{z} =z \ket{z} $
holds; $ z $ is proportional to complex amplitude of the classical
electromagnetic field obtained as the solution of Maxwell equation.
\refmark {\Loudon}
The variation equation now becomes
$$
\dot z + i \omega_0 z = -iE e^{-i\omega t},
\eqn\Cfive
$$
or, using the polar form $z=r e^{i\theta}$,
$$
\dot r = -E \sin(\theta + \omega t),~
r ( \dot \theta + \omega_0 ) = - E \cos(\theta + \omega t).
\eqn\Csix
$$
In a similar manner to the previous sections,
one can also have a \lq\lq resonance solution"
with period $ T= {2\pi \over \omega} $
which is given as
$$
r = r_0,~\theta=- \omega t ,
\eqn\Cseven
$$
where the relation
$$
r_0 = \left\vert { E \over \omega + \omega_0 } \right\vert
\eqn\Ceightx
$$
defines the \lq\lq invariant surface".
The phase $ \Gamma $ is thus evaluated as
$$
\Gamma(C) = 2 \pi \hbar r_0^2.
\eqn\Cnine
$$
On the other hand, the phase $ \Delta $ becomes
$$
\Delta(C) = {2 \pi \over \omega}(\hbar \omega_0 r_0^2 - 2E\hbar r_0)
\eqn\Cten
$$
As in the previous two cases, the phase $ \Gamma $ depends
only on the invariant surfaces and the Hamiltonian phase does not
possess with such a property.

\chapter{Possible Detection of the Geometric Phases}

We shall examine possible experimental detection of the geometric phases.
We first consider a general
setting to detect the phase with the aid of interference by using the
"particle beam" by which the coherent state is conveyed.

We suppose an interference apparatus consisting of two "arms".
(see Fig.1). Consider the incident beam in which the coherent state is
initially in the $ \ket{z} $ and it is splitted into
two beams running along two "arms". At the initial junction point
(the point A in the Fig.1),
the state is set to be in the state $ \ket{z} $.
The interference can occur in the following manner:
The state in one beam is set to be in the same state
as the initial one $ \ket{z} $, whereas the state in the other
beam is arranged such that the magnetic field or pseudo-magnetic
field is applied on this beam; hence,
after the time interval $ T $, it becomes
$ U(T)\ket{z} $. Thus if one considers the recombination at the
final junction point (the point B in the Fig.1),
the interference may be given by the superposition:
$$
\ket{\psi}={1\over 2}( \ket{z} + U(T)\ket{z})
$$
Actually, the interference can be observed
by the overlap $ \braket{\psi}{\psi} $, which turns out to be
$$
\braket{\psi}{\psi} = {1\over 4}(2+ \bracket{z}{U(T)}{z}
          +  \bracket{z}{U^{\dagger}(T)}{z})
$$
The cross term gives nothing but the propagator for
a round trip from $ z $  to $ z $, hence
$$ \braket{\psi}{\psi} =  {1\over 2}(1 + \cos\Phi(C)) =
                    \cos^2{\Phi(C) \over 2}
$$
Here we have two problems: the first is the problem of
"coincidence", namely, the time interval $ T $
should match the frequency $ \omega $ appearing in the
oscillatory magnetic or pseudo-magnetic field.
This suggests that during the travel of the beam along
the one arm whose length is taken to be $ L = cT $,
($ c $ means the beam velocity) the spin or pseudo-spin
figures the closed loop in the
Bloch sphere or pseudo-sphere in the complex $ z $ space.
Having assumed that this condition is satisfied,
we expect the effect of the interference due to the phase $ \Gamma $.

The second point to be mentioned is the problem concerning how
one can separate the geometric part
$ \Gamma $ from the total phase $ \Phi $. This may be possible,
if  one takes account of
the characteristics of the resonance solutions; namely,
if we choose the parameters $ (B,B_0, \omega) $
or $ (\kappa, \omega, \omega_0) $ such that the dynamical term $
\Delta $ vanishes, we can extract the effect which comes
only from the geometric part $ \Gamma $ alone.
On the basis of this general argument, we examine the
condition for which the phase $ \Gamma $ vanishes.
In order to see this, we consider the three different
cases separately.

(1): For the case of SU(2) CS, the condition
for which $ \Delta $ vanishes reads
$$
\cot \theta_0 ={ B_0 \over B },
\eqn\Yfour
$$
By combining this with the relation of "invariant surface" ~\Bten~,
we get
$$
\omega = -{\mu(B_0^2 +B^2) \over \hbar B}.
\eqn\Yfive
$$
The phase $ \Gamma $ thus becomes
$$   \Gamma(C) = 2J\pi(1 - {B_0 \over \sqrt{B_0^2 + B^2}}).
\eqn\Ysix
$$
As to the actual setting of experiment, this may be realized
by the particle beam consisting of particles of spin $ J $.
One of the splitted two beams is prescribed to be
placed in the magnetic field that oscillates sinusoidally.

(2):  For the case of the Lorentz coherent state, the
condition for which the dynamical phase vanishes is given by
$$
\coth \tau_0 =  -{\kappa \over \omega_0},
\eqn\Yseven
$$
where the inequality $ \vert {\kappa \over \omega_0} \vert > 1 $
should be satisfied. If we combine this with the relation
between the angle
$ \theta_0 $ and $ (\kappa, \tau, \omega) $, we get
$$
4\kappa^2 = (2\omega_0 +\omega)\omega_0.
$$
The phase $ \Gamma $ is calculated to be
$$   \Gamma(C) = 2k\pi\hbar({\vert \kappa \vert \over
      \sqrt{\kappa^2 - \omega_0^2}} - 1).
\eqn\Yeight
$$
In this case, the particle beam can be taken as the
coherent light (laser) beam; the coherent state is realized
by the squeezed state, which may be prepared appropriately.
If one of two splitted beams is controlled by pumping which
oscillates sinusoidally, we can expect the interference
pattern due to the geometric phase according to the
general formula.

(3): In a very similar manner, we can also arrange the
experiment for the case of the canonical coherent state,
for which the condition for $\Delta(C)$ to vanish is given by
$$
\omega = -{1 \over 2} \omega_0.
\eqn\Ynine
$$
The phase $ \Gamma $ is given by
$$
\Gamma(C)= {8\pi \over \omega^2}
           \vert E \vert^2  .
\eqn\Yten
$$
In this case, the experimental demonstration may also be
carried out by using the laser beam.

\chapter{Discussion and Summary}

As is seen from the result of the previous section,
the geometric phases nicely match the change of external field;
typically, sinusoidal oscillation characterizing the resonant
Hamiltonian. For this case, there exists a simple path on the
generalized phase space of CS (which is called the
\lq\lq resonance solution") when the external
parameters satisfy a certain condition(\lq\lq invariant surface").
In this way,  the geometric phase depends
only on the \lq\lq invariant surface". This feature ~enables
us to arrange the experiment such that the effect of the
phase is discriminated from the dynamical(or Hamiltonian)
phase which depends on the explicit form of the Hamiltonian.

 From the point of view of differential geometry,
the appearance of the geometric phases can be
regarded as a manifestation of the "holonomy" in quantum mechanics.
\refmark{\Simon}
As we have seen, the appearance
of the geometric phase is relevant only for the case of
non-stationary problem, namely,
the time-dependent Hamiltonian. Indeed this is very
contrast to the situation of the time-independent Hamiltonian.
Here a remark is given for this point. First to be mentioned is that
in the case of stationary case, the quantity with which we are
primarily intested is the energy eigenstate.
Thus what is a connection between the energy eigenstate
and the phase  $ \Gamma $ ?
Suppose an isolated system that is placed in the constant
external field.  Then the expectation value of the
Hamiltonian is time-independent
and the motion of the parameter $Z$ that determines the phase $ \Gamma $
lies on the surface of constant energy: $  H(Z,Z^*)= E $.
 For this case, after a cyclic change the semiclassical
transition amplitude $ K_{sc} $ acquires the phase factor except for the
energy factor: $ \exp[i\Phi] = \exp[{\Gamma(C)\over \hbar}] $
For an isolated system $ K_{sc} $ should be single-valued with
respect to Z; namely,
$$ \exp[i\Gamma(C)]=1
\eqn\Xone
$$
This is a reminiscent of the singlevalued nature of the usual wave
function leading to Bohr-Sommerfeld quantization, thus
\Ref\KuraX{
H.~Kuratsuji,~Phys.Lett.~{\bf 108B}~(1982)~367.}
$$
\Gamma(C)= \noint_C \bra{Z}i\hbar{\partial \over \partial t}\ket{Z} dt
         = 2 \pi \hbar n (n:{\rm integer}).
\eqn\Xtwo
$$
We here demonstrate the above statement by using the simplest
model Hamiltonian:
$ \hat H = - \mu B \hat J_z $ and spin CS.
Then $ H(Z,Z^*)= \mu B J \cos \theta = E ( = \mu B  m ), $
since $\cos \theta = {m \over J}$.  Therefore
$$ \Gamma(C) = 2\pi J \hbar ( 1 - \cos \theta_0 )
             = 2\pi \hbar(J - m),
\eqn\Xthree
$$
On the other hand, we know $ m $ (the z-component of the spin) takes
quantized integer (or half integer value) ranging from $ -J $ to $ +J $.
Thus, if $ \Gamma $ is exponentiated, we have
the trivial result.  In this way, the phase $ \Gamma $
should be called a "non-holonomic phase", if it is used for the
stationary state. On the contrary, we have the "holonomical" phase
only for the case when we consider the non-stationary quantum state
for which the concept of energy eigenstates loses the meaning, that is,
we have the concept of quasi-energy at best.

Finally, we shall point out possible perspectives
on the utility of the geometric phase inspired from the resonant
Hamiltonian that has not been treated here.
The Hamiltonian in the external field considered here is generic
and there may be possible phenomena that can be described
by these simple model Hamiltonians: for example,
the Hamiltonian relating to the
Bogoluybov equation in the superfluid $ {\rm He^3} $
and similar models inspired from condensed matter
physics.

\ack

The authors would like to thank the members of Department of
physics, Ritsumeikan university during this work was carried out,
especially, Mr.M.Muzui for his help in preparing the figure.
One of the authors (H.K.) is grateful to professor H.Rauch
of the Atom-Institut, Austrian Universities at Vienna for kindly
inviting him to deliver a seminar about
the subject discussed in the present paper.

\Appendix{A}

We summarize some necessary formulas for
the Lorentz coherent states.
\refmark{\Per}
The discrete series generators $K_i(i= 1,2,3)$ of SU(1,1)
algebra satisfy  the followig commutation relations :
$$
[ \hat K_0, \hat K_1 ] = i\hat K_2,
{}~[ \hat K_1, \hat K_2] = -i\hat K_0,
{}~[ \hat K_2, \hat K_0] = i\hat K_1,
\eqn\apaone
$$
which can be written formally as
$ [ \hat K_l, \hat K_m ]=i\tilde \epsilon_{imn} \hat K_n, $
where the symbol $ \tilde\epsilon $ is the same
as the one appearing in the "pseudo" scalar product \Lnine.
These are the abbreviavion of the usual commutation relation
$$
[ \hat K_0, \hat K_\pm ] = \pm \hat K_\pm,
{}~
[ \hat K_-, \hat K_+ ] = 2 \hat K_0,
\eqn\apathree
$$
where $ \hat K_\pm = i(\hat K_1 \pm i \hat K_2 ) $
are raising and lowering operators of a SU(1,1) state.
The eigenvectors of $K_0$ are specified by (k,m):
$$
\hat K_0 \ket{k,k+m} = (k+m) \ket{k,m},
\eqn\apafive
$$
where k is a real number determined by the representation of
SU(1,1) algebra and m is a non-negative integer.
Specifically, $ \ket{0} \equiv \ket{k,m=0} $ becomes a
starting state vector from which the Lorentz coherent state
is constructed. If we use
$$
\ket{m} \equiv \ket{k,k+m}
= \left[ {\Gamma(2k) \over m! \Gamma(m+2k)} \right]^{1/2}
(\hat K+)^m \ket{0},
\eqn\apasix
$$
we get the explicit form of the CS
$$
\ket{z} = (1- \vert z \vert ^2)^k \sum_{m=0}^\infty
\left[ {\Gamma(m+2k) \over m! \Gamma (2k)} \right]^{1/2} z^m \ket{m}.
\eqn\apaseven
$$
Furthermore noting that~$ \ket{m}$'s are mutually orthonormal, we get
the overlap for the un-normalized CS;
$$
\braket{z_1}{z_2}
={ [(1-\vert z_1 \vert^2)(1-\vert z_2 \vert^2)]^k
  \over(1-z_1^* z_2)^{2k} }.
\eqn\apbone
$$

Next some matrix elements are given for the generators of SU(1,1)
algebra by following the the same procedure for the case of SU(2) CS.
\refmark{\KS}
The matrix elements for $\hat K_+$  is calculated as
follows: Differentiating \LtwoX ~with respect to $z_2$,  we have
$$
 {d \over dz_2}(1-\vert z_2 \vert^2)^{-k} \braket{z_1}{z_2}
=\bra{z_1}{d \over dz_2} e^{z_2 \hat K_+}\ket{0}
=(1-\vert z_2 \vert^2)^{-k} \bra{z_1} \hat K_+ \ket{z_2}.
\eqn\apbtwo
$$
If using \apbone, this leads to
$$
{ \bra{z_1} \hat K_+ \ket{z_2} \over \braket{z_1}{z_2} }
=  { 2 k z_1^* \over 1- z_1^* z_2 }.
\eqn\apbthree
$$
In a similar manner, we also get
$$
{ \bra {z_1} \hat K_- \ket{z_2} \over \braket{z_1}{z_2} }
= { 2 k z_2 \over 1-z_1^* z_2 }.
\eqn\apbfour
$$
To derive the matrix element of $\hat K_0$, it is convenient
to use use the formula:
$$
e^{-z \hat K_+} \hat K_0 e^{z \hat K_+}
\hfill\break
= \hat K_0 + z \hat K_+,
$$
which yields
$$
\bra{z_1} \hat K_0 \ket{z_2}
=(1-\vert z_2 \vert)^k \bra{z_1}e^{z_2 \hat K_+}
(\hat K_0+z_2 \hat K_+) \ket{0} \break
=k \braket{z_1}{z_2} + z_2 \bra{z_1} \hat K_+ \ket{z_2}.
$$
By using this together with \apbthree, we have
$$
{ \bra{z_1} \hat K_0 \ket{z_2} \over \braket{z_1}{z_2} }
= { k (1+z_1^* z_2) \over 1 - z_1^* z_2}.
\eqn\apbfive
$$

\Appendix{B}

We point out a possible connection between
the resonance solutions and the NMR (nuclear magnetic resonance).
The NMR has been used for the study of the adiabatic phase
in a different context (See for example,
\Ref\Cina{
J.A.Cina, Chem.Phys.Lett.{\bf 132}(1986)393.}).
First to be noted is that the semiclassical solution satisfying
$ \delta S = 0 $ gives the exact solution for the
CS solution for the Sch$\ddot{\rm o}$dinger equation.
Note that the CS does not change its form, since the Hamiltonian (3.3)
involves the generators linearly.
\refmark{\Per}
Therefore we consider the solution of Schr$\ddot{\rm o}$dinger equation.
Let us first remind of the basic point of the NMR briefly.
Consider that the time evolution is given by
$ \hat U'(t)=e^{-{i \over \hbar}
\hat H't}
(\hat H'\equiv -\mu{\bf B}''
{\bullet}{\hat J}) $

in the moving frame (F$'$) which
rotates in x-y plane with the angular velocity $ \omega $.
On the other hand, in the static frame, say F, we have
$ \hat U(t) \equiv e^{-{i \over \hbar} \int  \hat H(t) \,dt}
=e^{-i \omega \hat J_z t} \hat U'(t)$, where
$ B'' $ represents the effective magnetic field
in the moving frame
\Ref\Mess{
A.~Messiah "M$\acute {\rm e}$canique Quantique"
{}~(Dunod,~Paris,~1958)}
which is given by
$$
{\bf B}''
={\bf B}'+{\hbar \over \mu}{\bf \omega}
=(B_0,0,B)+(0,0,{\hbar \omega \over \mu})
=(B_0,0,B+{\hbar \omega \over\mu})
\eqn\Yone
$$
Note that the operators in both frames are written
in the Schr$\ddot {\rm o}$dinger picture for each frame.
The spin variables in both frames satisfy the equation of motion.
$$
{d' \over dt}{\bf S}
={d \over dt}{\bf S}'
={\mu \over \hbar} {\bf S}' \times {\bf B}'' ,
\eqn\Ytwo
$$
which shows that ${\bf S}'$ makes a precession about the
magnetic field ${\bf B}''$, namely, this means that the spin nutates
in the static frame.
\Ref\Bloem{
N.~Bloembergen, "Nuclear Magnetic Relaxation"
{}~(W.A.Benjamin,~New York, ~1961);I.~I.~Rabi,~N.~F.~Ramsey and
J.~Schwinger,~Rev.~Mod.~Phys.{\bf 26}
(1954),167}
If we have  ${\bf S}'$//$ {\bf B}''$
at $t=0$, it follows that ${\bf S}'$ ( ${\bf S}$ as seen from F$'$ )
becomes constant in the moving frame $F'$,
$$
{\bf S}'(t)={\bf S}(t=0)=(J\sin \theta_0,0,-J\cos \theta_0),
\eqn\Ythree
$$
where
$$
\cot \theta_0=
-\left({B \over B_0} +{\hbar \omega \over \mu B_0} \right).
\eqn\Yfour
$$
Furtheremore, this shows that in the frame F,
${\bf S}$ purely rotates with the same frequency $\omega$ as
the magnetic field in the x-y plane:
$$
{\bf S}(t)=(J\sin \theta \cos \omega t, J\sin\theta\sin\omega t,
 -J\cos \theta)
$$
If camparison is made with \Ytwo, this leads to
$$
\phi=\omega t.
\eqn\Yfive
$$
\Yfour~and \Yfive~yield exactly the same conditions for $(\theta,\phi)$
as \Bnine ~and \Bten ~for "resonance" solution and "invariant surface".
In particular, if the condition
$$
\omega=\omega_0\equiv -{\mu B \over \hbar}
$$
satifies, the magnetic resonance occurs.
Finally, the similar argument may be applied for
the resonant Hamiltonian for the Lorentz coherent states,
which can be simply done by replacing the spin and  the magnetic
field by the "pseudo-spin" and  "pseudo-magnetic field".

\refout
\bye